\documentclass[%
    reprint,
    amssymb,
    amsmath,
    aip,
    prapplied,
]{revtex4-2}

\usepackage[english]{babel}
\usepackage{amssymb}
\usepackage{graphicx}
\usepackage{docs}
\usepackage{bm}
\usepackage{braket}
\usepackage[colorlinks=true,linkcolor=blue,citecolor=blue,urlcolor=blue]{hyperref}
\usepackage{siunitx}
\DeclareSIUnit\amagat{amg}
\DeclareSIUnit\molar{M}

\usepackage{color}
\usepackage{ulem}

\addto\extrasenglish{%

}

\expandafter\ifx\csname package@font\endcsname\relax\else
\expandafter\expandafter
\expandafter\usepackage
\expandafter\expandafter
\expandafter{\csname package@font\endcsname}
\fi

\makeatletter
\def\@email#1#2{%
 \endgroup
 \patchcmd{\titleblock@produce}
  {\frontmatter@RRAPformat}
  {\frontmatter@RRAPformat{\produce@RRAP{\phantom{*}\!\!#1\href{mailto:#2}{#2}
  }}\frontmatter@RRAPformat}
  {}{}
}%
\makeatother

\newcommand{\myaffiliation}{\affiliation}%\address}

\newcommand{\ICFO}
{\myaffiliation{ICFO -- Institut de Ci\`encies Fot\`oniques, The Barcelona Institute of Science and Technology, 08860 Castelldefels (Barcelona), Spain}}

\newcommand{\ICREA}{\myaffiliation{ICREA -- Instituci\'{o} Catalana de Recerca i Estudis Avan{\c{c}}ats, 08010 Barcelona, Spain}}

\begin{document}

\title{Local, mm-scale $^1$H magnetic resonance imaging using atomic vapors}

\author{Sven Bodenstedt}
\ICFO

\author{Morgan W. Mitchell}
\ICFO
\ICREA

\author{Michael C.D. Tayler$^\ast$}
\ICFO
\email[$\ast$ Electronic mail: ]{michael.tayler@icfo.eu}

\date{\today}%

\begin{abstract}
    \textbf{Abstract:} We demonstrate a practical and sensitive low-field approach to magnetic resonance imaging (MRI) of \si{\micro\liter}-scale samples using an optically pumped magnetometer (OPM).  Using a simple setup where a commercial OPM is placed directly adjacent to a sample, without intermediary signal-pickup coils, we perform one- and two-dimensional imaging with a $>$\SI{1}{cm} field of view and sub-mm Fourier-limited resolution at \SI{10}{\micro\tesla}, near Earth's field.  This enables (i) high-throughput, low-field MRI of fluidic and tissue samples, and (ii) a quantitative benchmark of the sensitive volume surrounding the OPM.   
\end{abstract}

\maketitle

\section{Introduction}
\noindent
Optically pumped atomic-vapor magnetometers (OPMs) are among the most precise magnetic-field sensors available today, reaching noise floors below $\SI{10}{\femto\tesla\per\sqrt{\hertz}}$ even in compact, mobile, mm-scale devices\cite{kominis_subfemtotesla_2003,griffith_femtotesla_2010, mitchell_colloquium_2020,fabricant_how_2023}. 
When such sensitivity is combined with spatial resolution --- through multichannel arrays \cite{kim_parallel_2020}, gradient encoding, or both --- powerful methods for mapping low-frequency ($<\SI{1}{\mega\hertz}$) magnetic fields become possible. Applications under development include magnetoencephalography (MEG) \cite{brookes_magnetoencephalography_2022,aslam_quantum_2023}
and ultralow-field magnetic resonance imaging (ULF-MRI)\cite{savukov_mri_2009,savukov_anatomical_2013,savukov_non-cryogenic_2013,hori_magnetic_2022}, both of which are progressing toward clinical feasibility at anatomically relevant scales.

A practical advantage of OPMs is their operation without cryogenics. In contrast to superconducting quantum interference devices (SQUIDs) \cite{mcdermott_microtesla_2004,zotev_squid-based_2007,espy_squid-detected_2013, lee_squid-based_2019, kempf_13c_2024}, OPMs can be positioned directly next to samples at ambient temperature, improving near-field sensitivity despite a slightly higher intrinsic noise. This practicality has renewed activity in related areas like near-zero-field nuclear magnetic resonance (NMR) spectroscopy \cite{jiang_zero-_2021, barskiy_zero-_2025} and nuclear-spin polarimetry \cite{mouloudakis_real-time_2023,eills_live_2024}. 
Still, MEG and MRI differ fundamentally in how spatial information is encoded: MEG relies on solving an inverse problem governed by dipolar field decay, whereas MRI imprints spatial coordinates onto the phase angle of nuclear precession. 
As a result, ULF-MRI systems, whether OPM- or SQUID-detection based, are typically configured with flux-transformer pickup coils between the magnetometer and sample, or other strategies (e.g., remote detection) to suppress the sensor’s intrinsic spatial sensitivity\,\cite{hori_magnetic_2022,xu_magnetic_2006}.
Additionally, transformer coupling relaxes geometric constraints by letting the sensor sit in a magnetically quiet region, away from the prepolarization and gradient fields. The trade-off is that it adds noise and bandwidth limits, and fixes the detection pattern to that of the coil.

In this work, we investigate the opposite detection strategy: placing the OPM directly next to the sample, eliminating the transformer and coupling the sensor to nuclear magnetization without intermediate elements.  This maintains conventional contrast mapping abilities, while also exposing the sensor’s intrinsic spatial sensitivity.
We begin by demonstrating position-encoded NMR signals [one-dimensional (1D) MRI] at \SI{10}{\micro\tesla} with sub-mm resolution and a field of view of up to \SI{1}{cm}.   
This is extended to $T_1$ relaxometry mapping of mm-scale tissue samples from fruits and vegetables, and then to two-dimensional (2D) MRI.  We discuss how this form of MRI may also provide a basis for sensor placement optimization and forward-calibration models relevant to MEG.

\section{Description of OPM-based MRI system} 

\begin{figure}[b]
    \centering
    \includegraphics[width=\columnwidth]{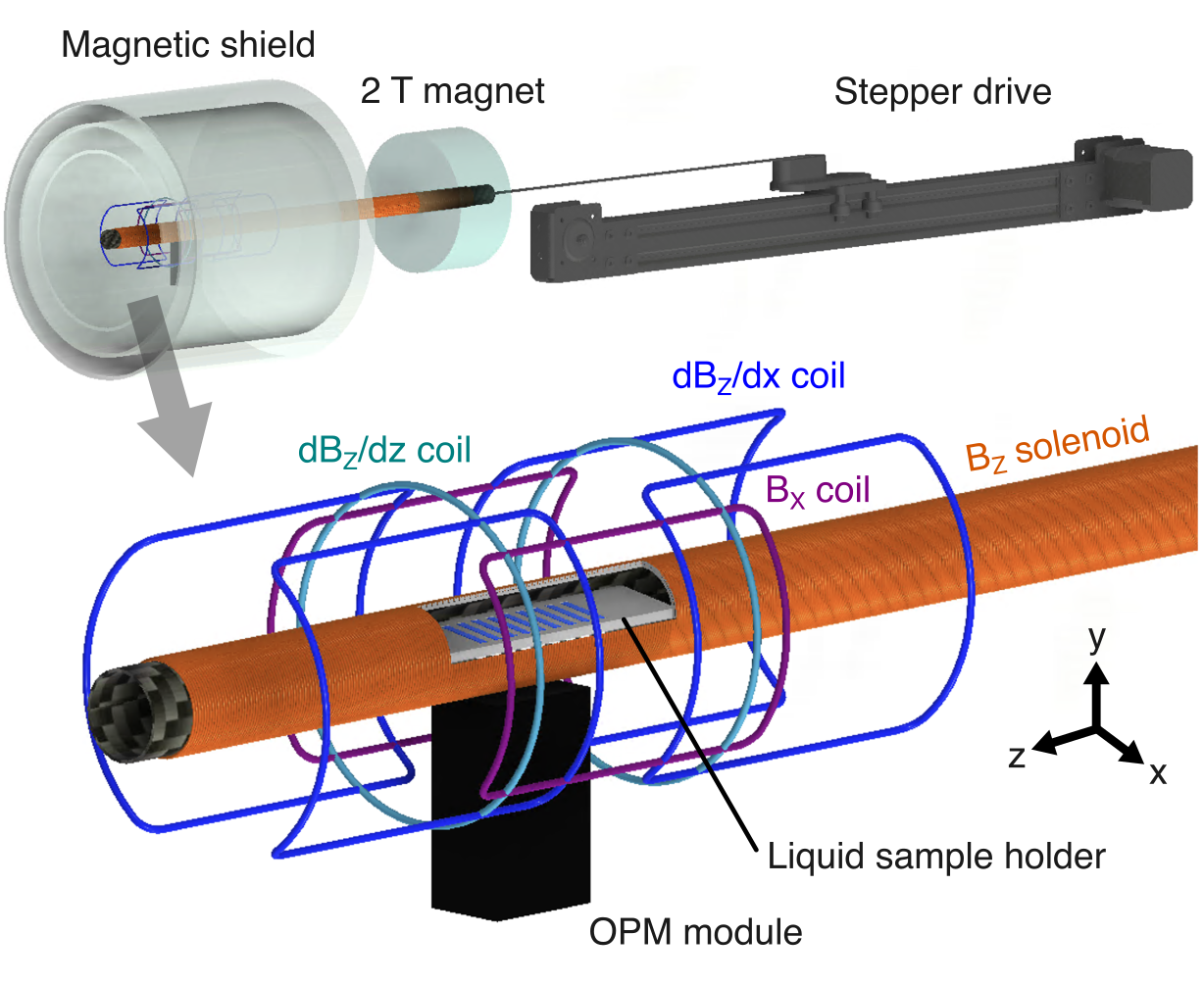}
    \caption{Experimental setup for ULF-MRI (to scale). The stepper drive precisely shuttles the sample back and forth between the external strong magnet for nuclear spin prepolarization and the shielded OPM for detection.}
    \label{fig:fig1}
\end{figure}

\subsection{Overview of components, sample preparation}

The overall system, illustrated in \autoref{fig:fig1}, was designed to maximize signal-to-noise ratios under the constraint that most commercially available OPMs operate in background fields only up to a few $\mu$T. 
To reconcile this requirement against the need to polarize nuclear spins in a high field, we used a shuttle architecture to perform polarization and detection at separate locations;
a stepper-motor belt drive system (OpenBuilds mini-V linear actuator, step precision 0.05 mm) mechanically transported the imaging samples between a \SI{2}{T} low-homogeneity NdFeB magnet\cite{tayler_low-cost_2017} (bore dimensions: 15 mm diameter, 120 mm length, field uniformity $>$95\%) and a magnetically shielded region containing a commercial OPM sensor (Hamamatsu Photonics, 4-channel \textsuperscript{87}Rb OPM, pre-production prototype version; sensor dimensions $l_x=$ 13 mm, $l_y=$ 31 mm, $l_z=$ 19 mm, located in a Twinleaf model MS-2 magnetic shield\cite{twinleafweb}, dimensions 60\,cm length and 30\,cm diameter, residual background field approximately \SI{10}{\nano\tesla}). 
The approximately \SI{0.4}{m} distance between these two regions took around \SI{500}{ms} under a low-jerk belt velocity profile and demonstrated precise, repeatable positioning.

A solenoid coil (${\rm d}B_z/{\rm d}I=$ 9.52\,mT/A) was wound around the entire length of the shuttling tube (\SI{14}{mm}outer diameter, \SI{0.5}{mm} wall carbon-fiber) to provide the main detection field $B_z$.
Additional coils, all centered with respect to the midpoint of the shield, were supplied to spatially encode the images: (i) a transverse dc field $B_x$ (${\rm d}B_x/{\rm d}I=$ \SI{1}{\milli\tesla/A}), (ii) a linear gradient field $G_z={\rm d}B_z/{\rm d}z$ (anti-Helmholtz geometry, \SI{50}{mm} diameter, ${\rm d}G_z/{\rm d}I \approx$ \SI{5.0}{n\tesla/mm/mA}), and (iii) a linear gradient $G_x={\rm d}B_z/{\rm d}x$, (double-saddle geometry \cite{siebold_gradient_1990}, \SI{50}{mm} diameter, ${\rm d}G_x/{\rm d}I \approx$ \SI{8.5}{n\tesla/mm/mA}). These coils were integrated on a single 3D-printed bobbin surrounding the solenoid and shuttling tube and were hand-wound with enameled copper wire.

The samples were held within cylindrical segments, stereolithographically printed from ``Rigid 10k'' UV-curing photopolymer material (3D printer: FormLabs Form 3, print resolution \SI{25}{\micro\meter}).  The cylinder diameter, 12 mm, matched the inner diameter of the $B_z$ solenoid minus a clearance of \SI{0.5}{mm}, while the height was around \SI{8}{mm} and the length around \SI{50}{mm}.  Cavity shapes such as circular holes or slots were incorporated into the structures to a depth up to \SI{7}{mm} from the flat surface as shown in the Figures.  For instance in \autoref{fig:1DMRI}a, smallest cavities were \SI{8}{mm} wide along $x$ and \SI{1}{mm} thick along $z$, thus each contained a volume of \SI{50}{mm\cubed} $\equiv$ \SI{50}{\micro\liter}.  The cavities were filled by adding liquid sample via a graduated pipette, or by inserting cut-to-size solid samples with tweezers.  A cover slide (Parafilm\textsuperscript{\textregistered} M) was added to seal the flat surface and prevent evaporation.

\subsection{Optical magnetometry}

The commercial OPM sensor represented by the `black box' in \autoref{fig:fig1} was positioned immediately below the solenoid coil. 
Here,
with approximately \SI{2}{mm} $y$-axis standoff distance between the sample and the sensor package edge it experienced only around 0.5\% of the \SI{10}{\micro\tesla} imaging field, to stay comfortably within the near-zero-field operating regime\cite{Bodenstedt2021natcomm}.  
The sensor was known to employ a single continuous-wave, circularly polarized beam of laser light along $z$, which simultaneously pumped and probed the D spectroscopic transition of a \textsuperscript{87}Rb vapor \cite{shah_fully_2018} confined to a mm-scale `vapor cell' therein.  The vapor cell was known to be centered 6-\SI{7}{mm} inside the sensor package along $y$; no other specific details were known about the vapor cell size or composition.  The $x$- and $y$-axis field components were output from the sensor via lock-in quadrature signals of the Rb Hanle resonance, and streamed out as a time-varying voltage from the OPM control unit.

Assuming calibrated coils, the typical set up procedure for the OPM involved setting desired $B_z$ and $G_z$ fields in the solenoid and then applying a low-amplitude (\SI{10}{\pico\tesla\textsubscript{rms}}) test current in the $y$-axis coil near the expected \textsuperscript{1}H Larmor frequency.  The OPM response band was shifted to this frequency by manually adjusting (i) the laser wavelength, and (ii) the local $z$ bias field at the vapor cell (via a $z$ coil located inside the sensor package) to around $\SI{425}{Hz}/\gamma_{\rm Rb}$, to maximize the $y$-axis signal response.  
After setup, the test current was turned off.  

To quantify noise, the amplitude spectral density (ASD) profile of the $y$-axis signal was used and was defined as the Welch-averaged median ASD over the \SI{400}{Hz}–\SI{450}{Hz} band.  We quote the ASD in field units after normalization against the ASD peak with the known test-signal current stated above.
Using this in situ calibration, we measured a noise floor of 25-\SI{30}{\femto\tesla\per\sqrt\hertz} with $B_z=\SI{10}{\micro\tesla}$ at the sample and a compensated near-zero field at the OPM vapor cell. 
This is comparable with OPMs from other manufacturers, as used in past ULF-NMR studies\cite{lee_-situ_2019,blanchard_zero-_2020}.

\subsection{MRI pulse sequences and data processing} \label{sec:methodsMRI}

\begin{figure*}[t]
    \centering
    \includegraphics[width=0.95\textwidth]{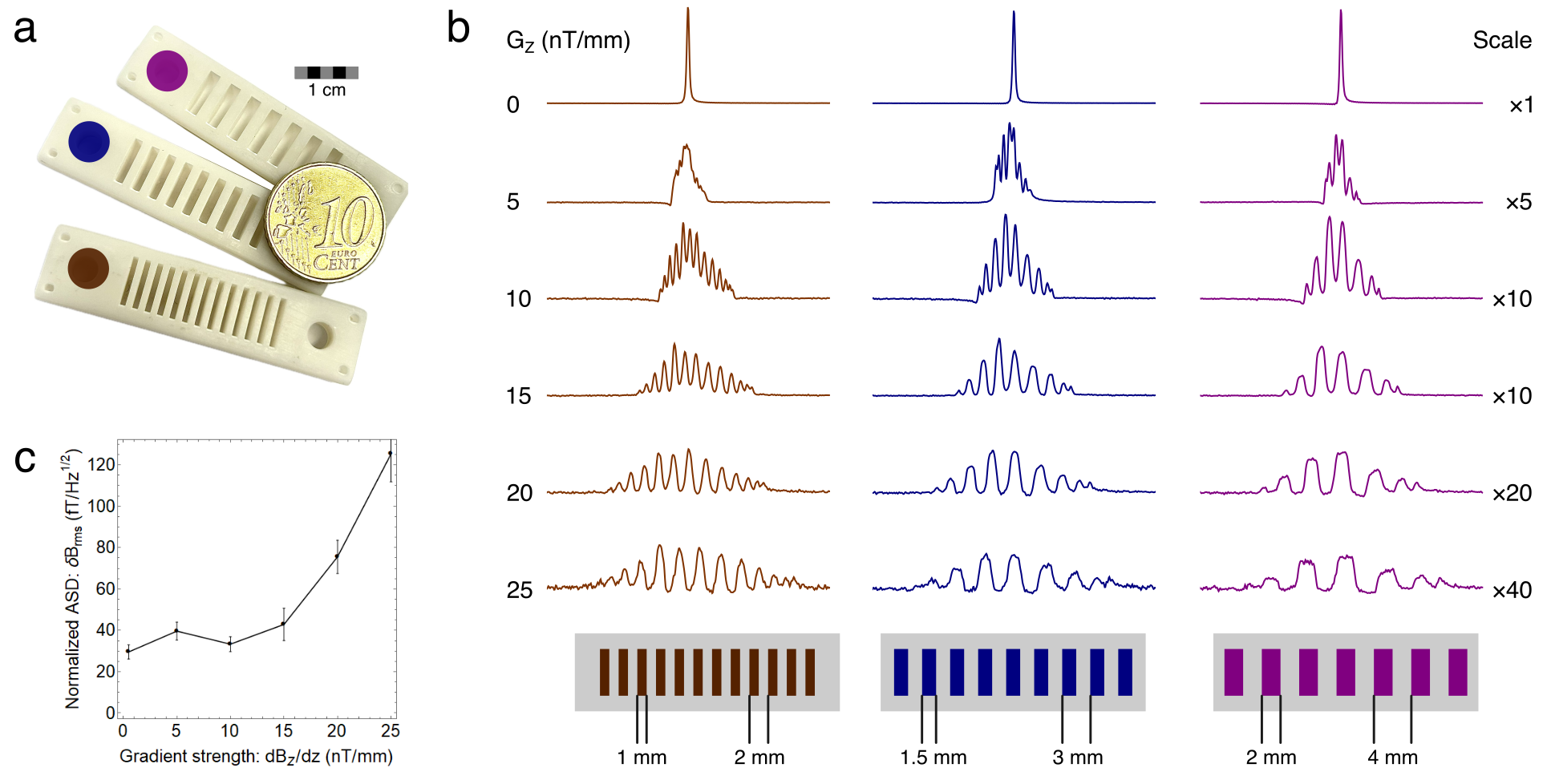}
    \caption{Fixed gradient 1D \textsuperscript{1}H NMR images: 
    (a) alternating water-cavity/wall structures used as test phantoms; 
    (b) 1D images measured at \SI{10}{\micro\tesla} using a commercial OPM.  
    The three columns correspond to cavity pitches of (left) \SI{2}{\milli\meter}, (middle) \SI{3}{\milli\meter} and (right) \SI{4}{\milli\meter}, while rows show spectra acquired for different gradient strengths $G_z$. Each spectrum has a width of \SI{40}{\hertz} and is the sum of 16 transients;
    (c) variation in OPM sensitivity with $G_z$, which above \SI{15}{\nano\tesla/\milli\meter} increases mostly due to uncompensated dc fields.
    }
    \label{fig:1DMRI}
\end{figure*}

After ex situ polarization and return of the sample to the magnetic shield via the shuttle, a dc $(\pi/2)_x$ pulse was used to initiate \textsuperscript{1}H precession about the $B_z$ axis.

The simplest form of MRI -- spatial resolution along $z$ -- was achieved by measuring the OPM signal under finite values of $B_z$ and $G_z$ so the nuclear Larmor frequency varied according to $\omega(\bm{r}) = \gamma_{\rm H} (B_z + {\bm{G}\cdot\bm{r}})$; the output signal from the $y$-axis channel was thus expected to be proportional to 
$\int_{\bm{r}}{\rm d}\bm{r} a(\bm{r}) \rho(\bm{r})\cos[\omega(\bm{r}) t]$.  Here $\gamma_{\rm H}/{(2\pi)}\approx$ \SI{42.58}{Hz/\micro\tesla} is the value of the \textsuperscript{1}H gyromagnetic ratio, $\bm{G}$ is the gradient field, $\rho$ is the spin density, $a(\bm{r})$ is the OPM's position-dependent sensitivity and $\bm{r} = (x,y,z)$ is the position vector of spins in the sample with respect to the coil center.  

This expression was simplified given the relatively weak gradient strength, where the contribution of $\bm{G}$ to $\omega$ is below $1\%$ of $\gamma_{\rm H} B_z$. We thus ignore the concomitant (nonsecular) components and assume $\omega(\bm{r}) = \gamma_{\rm H} (B_z +G_zz)$.

To apply the field pulse and synchronously digitize and store the $y$-axis OPM signal during the subsequent precession period, we used an NMRduino spectrometer\cite{tayler_nmrduino_2024} (16-bit sampling, approximately \SI{150}{\micro\volt\per bit}, sampling rate $f_s=$ \SI{5}{kHz}).  The NMRduino stored the data on a computer for processing and plotting.  Fourier transformation of the signals produced the one-dimensional images shown in \autoref{fig:1DMRI}b.  Typical digital resolution along $z$ was $\Delta_z = 2\pi f_s / (n_s G_z \gamma_{\rm H})$ or $\Delta_z=$ \SI{1.1}{mm} using $G_z=$ \SI{6}{nT/mm}.
  
For 2D image encoding (see \autoref{fig:fig4}) the above method of acquiring signals under a finite $G_z$ was combined with standard phase encoding in $x$.  Instead of acquiring the OPM signal immediately after the $\pi/2$ excitation pulse, we first applied a spin-echo sequence ($\tau-\pi_x-\tau$), where each half-echo delay ($\tau=$ \SI{90}{ms}) contained a rectangular gradient pulse of length $\tau_G<\tau$ and amplitude $\pm G_x$ through the ${\rm d}B_z/{\rm d}x$ coil.  Opposite amplitudes $+G_x$ and $-G_x$ were applied during the two halves of the echo to impart a precession phase shift of $\phi(\bm{r}) = 2 \gamma_{\rm H} \tau_G G_x x$.  A series of signals was measured starting after echo for a uniform set of gradient pulse durations $n \times \delta \tau_G$ ($n \in \{-n_{\max}, -n_{\max}+1, \ldots, n_{\max}\} \subset \mathbb{Z}$), and a discrete Fourier transform (DFT) in $n$ yielded an image with an $x$-axis field of view ${\rm FOV}_x = \pi/(\gamma_{\rm H} G_x\delta \tau_G)$ and digital resolution $\Delta_x = {\rm FOV}_x/(2n_{\max}+1)$.  In practice, we used $n_{\max}=32$, $G_x=$ \SI{170}{nT/mm} and $\delta \tau_G=$ \SI{2.5}{ms}, leading to FOV$_x = \SI{59}{mm}$ and digital resolution \SI{0.9}{mm}.  After DFT both along $z$ and $x$ dimensions, the images were phase corrected up to linear order in position and plotted as the real component.

\section{One-dimensional ULF-MRI} \label{sec:1dmri}

To benchmark the sensitivity and the resolution of the ULF-MRI system, we recorded \textsuperscript{1}H precession signals at $B_z=\SI{10}{\micro\tesla}$ ($\omega/(2\pi) \approx $ \SI{425}{Hz}) for deionized water in sample holders comprising an alternating rectangular cavity/wall/cavity structures along $z$.  Three different pitches of cavity were used, 
2 mm (\SI{1}{mm} thick, \SI{40}{\micro L} volume),  
3 mm (\SI{1.5}{mm} thick, \SI{60}{\micro L} volume), 
4 mm (\SI{2}{mm} thick, \SI{80}{\micro L} volume), 
resulting in 13, 9 and 7 cavities per sample holder, respectively (\autoref{fig:1DMRI}a).

In \autoref{fig:1DMRI}b, we compare the signals across gradient strengths from $G_z=$ \SI{0}{nT/mm} to $G_z=$ \SI{25}{nT/mm}, where the plotted spectra show the real part of the signal after Fourier transformation and a simple linear phase correction.  Even with $G_z$ on the order of tens of nT/mm, sub-mm image resolution is readily achievable and the spectral peaks from at least 5 adjacent liquid-filled cavities can be resolved without difficulty in a single 1D measurement.  At the highest gradient strengths, the distinct squared-off profile of peaks confirms a regime where the position-dependent gradient shift is large compared with the natural line width, or full-width at half-maximum height $1/(\pi T_2)$.  The near-constant peak-to-peak frequency difference within each spectrum confirms good linearity of the gradients, except at the ends of the sample. 

Higher gradient strengths increase the spatial resolution at the cost of signal-to-noise ratio.  Aside from the spreading effect of the gradient on the nuclear precession frequency, the noise of the OPM sensor also degrades as $G_z$ increases.  A possible reason is the shortening of the effective transverse coherence time $T_{2,\rm atom}^*$ of the sensor atoms, which increases the atomic shot noise, $\delta B_{\rm sn}\propto 1/\sqrt{T_{2,\rm atom}^*}$.
However, this is not supported as a major mechanism by the experimental data.  The decoherence time $1/T_{2,\rm atom}^*$ should in principle scale quadratically with gradient amplitude and, moreover, weaker gradients should be more damaging: for instance in Ref \cite{li_comprehensive_2024}, a gradient of only \SI{2}{nT/mm} along the beam was enough to reduce sensitivity by a factor of 3.  
In our case, the commercial OPM’s noise floor in the 
\SIrange{400}{450}{Hz} detection band 
remained flat below \SI{50}{\femto\tesla/\sqrt{Hz}} for $G_z$ up to \SI{15}{\nano\tesla/\milli\meter} (see \autoref{fig:1DMRI}c).
This measurement directly characterizes the OPM noise floor under finite-$G_z$ imaging conditions.

The degradation observed above \(G_z\approx \SI{15}{nT/mm}\) is attributed mainly to uncompensated dc offset fields from the gradient coil, beyond the available compensation range of the OPM.
The vapor cell was not exactly centered on the $G_z$ coil and thus the OPM's inbuilt shim coils were unable -- unlike for $G_z$ below $\SI{15}{\nano\tesla/\milli\meter}$ -- to null the offset for best sensitivity.  Extending the range of dc field compensation should overcome this limitation.

Another quantitative feature is the amplitude variation across each 1D image, which for high $G_z$ is consistent with the distance decay of the nuclear dipolar field: $a(z) \propto (1+(z/d_0)^2)^{-3/2}$.  The expression given here is the inverse-cube distance between the alkali and nuclear spins assuming a closest approach $d_0$, i.e., the inverse-cube magnitude of the interspin vector $(0,d_0,z)$.
We note the curve $a(z)$ has zero slope at $z=0$ and a half-maximum intensity $a(z') = a(0)/2$ at points $z' = \pm d_{0}\,\sqrt{2^{2/3}-1} \approx \pm0.76\, d_0$.  
It is from the latter that we define an ``above-50\%-amplitude'' field of view: ${\rm FOV}_{z,50} = 2z' \approx 1.5 d_0$ or around \SI{12}{mm} for $d_0 \approx$ \SI{8}{mm}, which is consistent with the one-dimensional image given the known sample dimensions.  An ``above-90\%-amplitude'' field of view is determined similarly from the full width at 90\% of maximum, or ${\rm FOV}_{z,90}  \approx 0.54 \,d_0$, around \SI{4}{mm} for $d_0 \approx$ \SI{8}{mm}, within which signal pickup should be uniform within 10\%.

\section{Spatially resolved relaxometry of mm-scale tissue specimens} \label{sec:tissueT1}

\begin{figure}
    \centering
    \includegraphics[width=0.99\columnwidth]{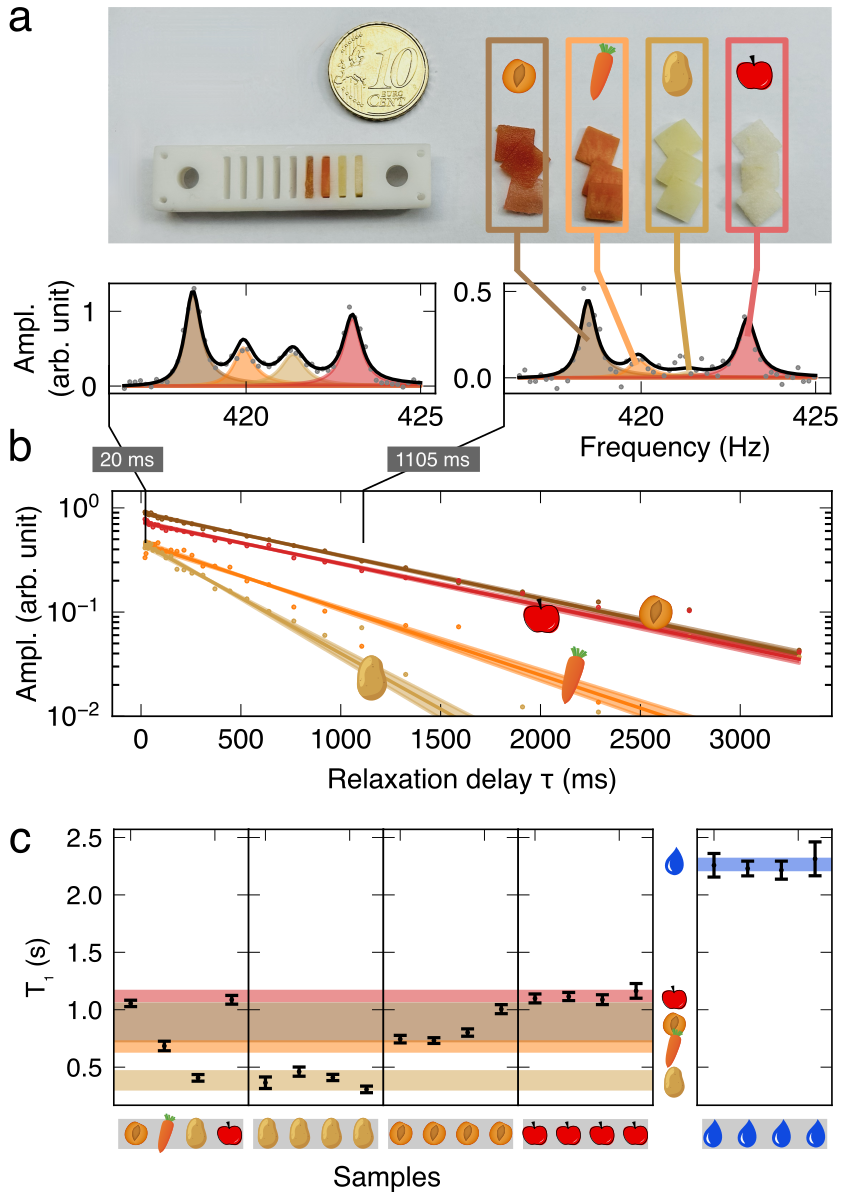}
    \caption{OPM-detected, relaxation-weighted 1D \textsuperscript{1}H MRI of vegan tissue slices ($6\times6\times\SI{1}{mm^3}$) at \SI{10}{\micro\tesla}.
    (a) Schematic of the sample holder.
    Plots show the real part of 1D spectra at relaxation delays $\tau$ of 20 and \SI{1105}{ms} after shuttling.  Peaks are fit with complex Lorentzians; black curve shows the total fit.
    (b) Fitted peak amplitudes vs.\ relaxation delay (semilog scale). Solid lines indicate single-exponential fits to $a_0 \exp(-\tau/T_1)$.
    (c) Fitted $T_1$ values for each sample configuration (peach + carrot + potato + apple, potato, peach, apple, H\textsubscript{2}O as control). Error bars denote standard error in $T_1$ fits.
    }
   \label{fig:T1}
\end{figure}

\begin{figure*}
    \centering
    \includegraphics[width=0.82\textwidth]{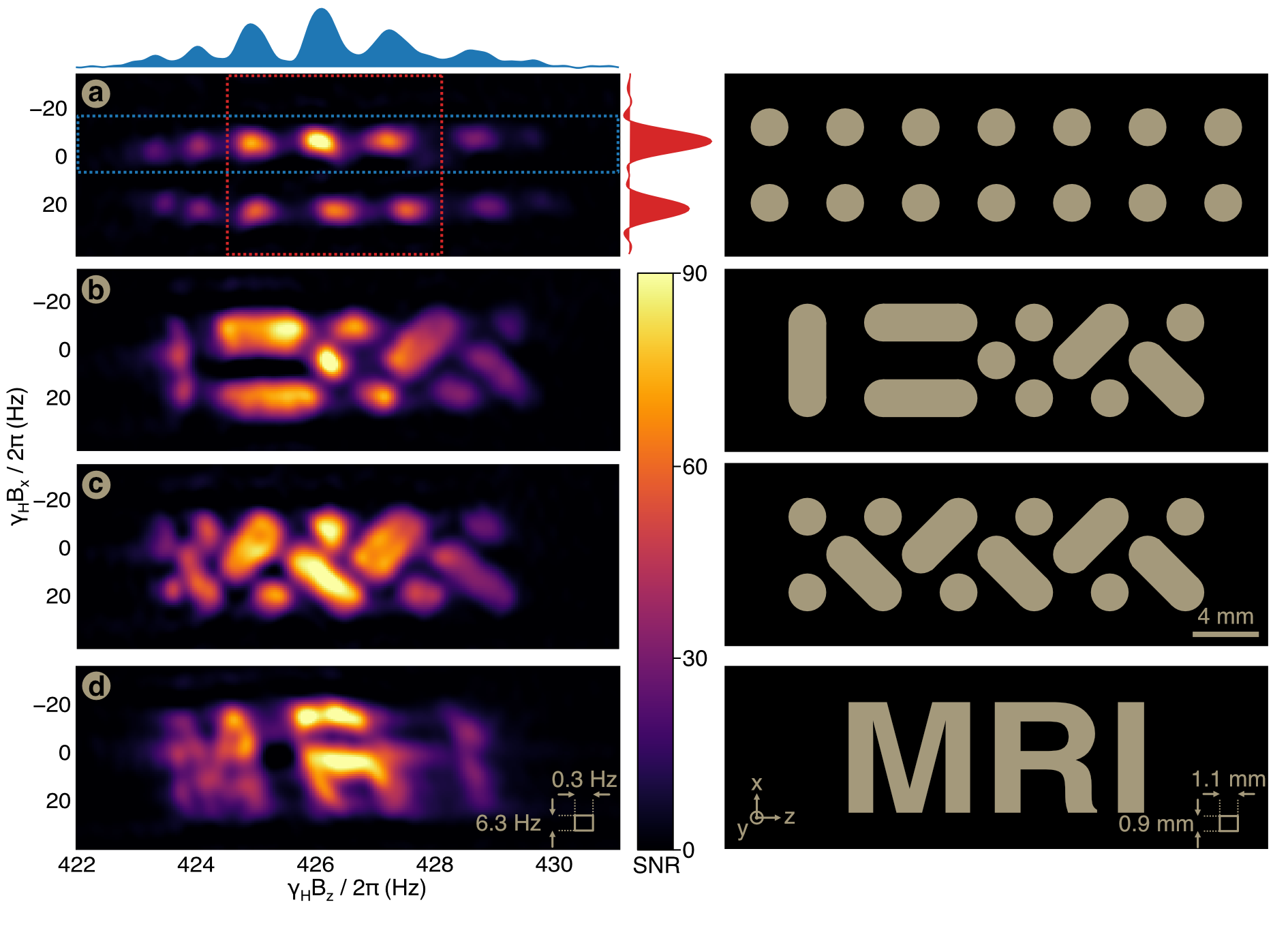}
    \caption{OPM-detected
    2D ULF-MRI at $B_z=$ \SI{10}{\micro\tesla}. 
    (a-d) Images of \textsuperscript{1}H\textsubscript{2}O in four 3D-printed phantoms, internal structures depicted to the right.
    The images were acquired using a spin-echo sequence with frequency encoding along the horizontal axis ($z$) and 65-step phase encoding along the vertical axis ($x$). 
    Images show the real component after 2D discrete Fourier transformation and linear phase correction. 
    The color scale represents signal-to-noise ratio (SNR). 
    1D projections of image `a' onto $z$ and $x$ are shown to the top and right of the raster, respectively; the central peaking in all images is characteristic of the OPM's distance-dependent sensitivity. 
    All images present a Fourier-limited in-plane resolution of $0.9\times\SI{1.1}{\milli\meter}$.
    }
    \label{fig:fig4}
\end{figure*}

The motivation for low-field MRI using either inductive pickup coils or OPMs is commonly framed relative to established high-field $(>\SI{1}{\tesla})$ MRI. 
Low-field systems trade sensitivity for simpler infrastructure, lower cost, improved materials compatibility, and access to alternative $T_1$ and $T_2$ relaxation contrast mechanisms, and in anatomic imaging these features often serve as a practical complementary regime \cite{arnold_lowfield_2023, webb_five_2023}.
In this light, we applied the 1D imaging protocol to samples of increased complexity, including cavities containing millimeter-scale biological specimens. 
In this architecture, \textsuperscript{1}H $T_1$ relaxation provides quantitative contrast through a delay $\tau$ introduced between shuttling and excitation, which modulates the longitudinal magnetization prior to readout by a factor approximately equal to $\exp{(-\tau/T_1)}$. 
Sensitivity to $T_1$ contrast below \SI{100}{\micro\tesla} has previously been demonstrated using SQUID-detected ULF MRI \cite{lee_squiddetected_2005} and fast-field-cycling methods \cite{kolodziejski_markers_2024,koenig_field-cycling_1990}, where field-dependent relaxation enhances contrast relative to conventional clinical fields.

In \autoref{fig:T1}, we show $T_1$-weighted 1D imaging of four biological tissue slices -- apple, carrot, potato, and peach -- placed in adjacent cavities of the 3-\si{\milli\metre}-pitch holder, each cut to 6 $\times$ 6 $\times$ \SI{1}{mm^3}. 
Under the applied gradient, the specimens are resolved in frequency space and quantified within the same acquisition with minimal crosstalk.
As \autoref{fig:T1}b shows, the fitted amplitudes exhibit mono-exponential decays that yield distinct $T_1$ values for each tissue type. 
The contrast arises from intrinsic differences in water \textsuperscript{1}H mobility, with denser tissues exhibiting shorter $T_1$. 
The fitted values remain consistent across sample configurations, as shown in \autoref{fig:T1}c, indicating that the measured contrast reflects intrinsic relaxation.

\section{2D MRI with sub-mm spatial resolution}

Spatial resolution in two dimensions at $B_z=\SI{10}{\micro\tesla}$ was demonstrated using a phase-encoded version of the 1D MRI sequence. The method involved pulsed ${\rm d}B_z/ {\rm d} x$ gradients fields during a spin echo, as described in \autoref{sec:methodsMRI}.  
A separate set of water-filled sample holders was used, with cavity shapes varying in both the $x$ and $z$ directions: (a) a rectangular grid of cylindrical cavities; (b) and (c) cylindrical or slot cavities of different orientation and length, and; (d) a text phantom spelling `MRI'.  

The images shown in \autoref{fig:fig4} were reconstructed using a 2D fast Fourier transform applied to datasets comprising 65 phase-encoded 1D slices, giving Fourier resolutions \SI{0.9}{mm/pixel} in $x$ and \SI{1.1}{mm/pixel} in $z$.   This is comparable to the resolution of alternative OPMs.  
Despite the simple data processing being limited to a simple linear phase correction and excluding deconvolution, image shearing, or correction for the distance-dependent pickup response $a(z)$, the images faithfully reproduce the original cavity shapes.  

The main image artifacts are a slight horizontal shear and a localized anisotropic compression of the $z$ axis near $\omega/(2\pi) = \SI{424}{\hertz}$.  In principle, these can be mitigated by advanced postprocessing as mentioned above, or by refining the linearity of the gradient coils in hardware.  Already, however, the result is a promising step to applications because the central part of the image between 425 and \SI{428}{\hertz} displays almost no distortion and high signal-to-noise ratios ($>$60).  
This performance level should be diagnostically useful in screening, especially when combined with low-field relaxation contrast weightings.  More broadly, this also demonstrates a feasible route toward efficient spatial parallelization of ULF NMR.  
High-throughput has hitherto only been demonstrated using multiple samples and multiple sensors \cite{Andrews2025PNASNexus}; by contrast our work involves a single sensor to simultaneously detect multiple samples, each 10--\SI{15}{\micro\liter}.

\section{Discussion and Conclusions}

The 1D and 2D images obtained in this work extend a practical regime of ULF-MRI to non-anatomic, cm-scale samples and architectures compatible with parallelization. 
A main feature is the placement of the OPM directly adjacent to the sample, without intermediary pickup coils or flux transformers. In this geometry, the OPM maintains -- relative to these alternative approaches\cite{savukov_mri_2009,savukov_anatomical_2013,lee_squid-based_2019} -- a low tens-of-\si{fT\per\sqrt{Hz}} noise floor over the acquisition times used, enabling single-scan 1D images of \si{\micro\liter}-scale fluid-filled cavities and 2D images over a field of view of approximately \SI{1}{cm^2}.

The 2D imaging example provides a useful estimate of small-volume sensitivity.  The central cylindrical cavity in \autoref{fig:fig4}a has an internal liquid volume of approximately \SI{13}{\micro\liter} and gives a peak pixel SNR of about 90.  This corresponds to an estimated sensitivity of order \SI{0.1}{\micro\liter}/pixel/scan\(^{1/2}\) for SNR approximately 1, which should be understood as a pixel-based sensitivity estimate for the present geometry and reconstruction, rather than a measured voxel detection limit.

While the spatial resolution in our work already matches the existing state of the art in ULF MRI, it could be increased further by improvements in sensitivity. 
As in conventional MRI, spatial resolution and sensitivity are coupled variables: a stronger gradient distributes the signal over a wider region of phase space at the expense of SNR per pixel. 
Other off-the-shelf OPM modules have reached below \SI{10}{fT\per\sqrt{Hz}}, which would directly improve the sensitivity. 
Alternatively, faster shuttling and higher prepolarization fields would increase the detected magnetization. This is particularly relevant for tissue samples with short \(T_1\) values -- for reference, excised prostate tissue exhibits \textsuperscript{1}H \(T_1\) values on the order of \SI{0.2}{s} at \si{\micro\tesla} fields \cite{busch_measurements_2012}, which appears within reach for future iterations of the apparatus.

A full characterization of the OPM physics under imaging gradients is beyond the scope of the present work, but is a significant next step. The relevant factors include the alkali-vapor spin coherence time, \(T_{2,\rm Rb}\), and gradients at the vapor cell caused either by uncompensated magnetic fields or by effective gradients associated with vector ac Stark shifts during optical pumping \cite{Sulai2013-hn,Jimenez-Martinez2014-jv}. Studies of this type are ongoing and will be reported in a forthcoming paper.

The second aspect of the experimental geometry is exposure of the MRI signal to the spatial response function of the OPM.  
Beyond the proposed MRI analyses oriented towards liquid or tissue samples, the ability to measure this intrinsic spatial response may be useful in MEG and other imaging applications currently under development with OPMs, e.g., simultaneous MEG-MRI \cite{Vesanen2012, Ito2024}.
In MEG, the reconstruction of neural current maps from the signal of each head-mounted OPM sensor depends on a `forward model' comprising accurate knowledge of each sensor’s position, orientation and spatial sensitivity. 
In our experiment, the source geometry is known from the shape of the MRI phantom, and -- all other factors being equal -- the measured signal amplitude is proportional to the OPM pickup response.
Because the MRI measurements reflect quasi-static nuclear magnetization in a controlled geometry, whereas MEG involves dynamic current sources, differences in source characteristics must be considered before directly transferring these spatial response functions.

\section*{Acknowledgments}
The work described is funded by: 
the Spanish Ministry of Science MCIN with funding from European Union NextGenerationEU (PRTR-C17.I1) and 
by Generalitat de Catalunya ``Severo Ochoa'' Center of Excellence CEX2019-000910-S; 
the Spanish Ministry of Science projects 
MARICHAS (PID2021-126059OA-I00), 
SEE-13-MRI (CPP2022-009771),
AMAREI (CNS2023-144106)
plus 
RYC2022-035450-I and 
SAPONARIA (PID2021-123813NB-I00), funded by MCIN/AEI /10.13039/501100011033; 
Generalitat de Catalunya through the CERCA program;  
Ag\`{e}ncia de Gesti\'{o} d'Ajuts Universitaris i de Recerca Grant No.  
2021 FI\_B\_01039; 
Fundaci\'{o} Privada Cellex; 
Fundaci\'{o} Mir-Puig.
The authors thank Norihisa Kato, Takeshi Endo and Jordi Sobrino of Hamamatsu Photonics K.K. for loan of the OPM sensor.

\section*{Competing interests}
The authors declare no competing interests. 

\section*{Data availability}
Experimental data files supporting this work are openly accessible via the web at \url{https://doi.org/10.5281/zenodo.20411497}.

\bibliography{ZULFMRI2,references}

\end{document}